# TSIA: A Dataflow Model

Burkhard D. Steinmacher-Burow

*Postfach 1163, 73241 Wernau, Germany*
*burow@ifh.de*
*http://www.tsia.org*

February 18, 2000

The Task System and Item Architecture (TSIA) is a model for transparent application execution. In many real-world projects, a TSIA provides a simple application with a transparent reliable, distributed, heterogeneous, adaptive, dynamic, real-time, parallel, secure or other execution. TSIA is suitable for many applications, not just for the simple applications served to date.
This presentation shows that TSIA is a dataflow model - a long-standing model for transparent parallel execution. The advances to the dataflow model include a simple semantics, as well as support for input/output, for modifiable items and for other such effects.

## 1  Introduction

For a computer application, a transparent execution is one not visible to the application. Instead, the execution is managed by a system external to the application.

The Task System and Item Architecture (TSIA) is a model for transparent application execution. The name TSIA is used for both the model and for systems implementing the model. In many real-world projects, a TSIA provides a simple application with a transparent reliable, distributed, heterogeneous, adaptive, dynamic, real-time, parallel, secure or other execution [Dividing]. As shown in detail elsewhere, TSIA is suitable for many applications, not just for the simple applications served to date [TSIA].

Dataflow is a long-standing model for transparent parallel execution [Models]. This presentation shows that TSIA is a dataflow model.

TSIA and the dataflow model are introduced in sections 2 and 3. The simple state of an executing application is described in section 4.

In many previous dataflow models, the execution of a routine effectively begins by macroexpanding the body of the routine. A recent advance over macroexpansion called delegation is described in section 5. A resulting alternative implementation of routines is described in section 6. It allows for a simple application execution as described in section 7.

An application consists of instructions, data and other items. As described in section 7, the semantics of TSIA do not include any of the items. This is in contrast to many previous dataflow models.

Delegation is more general than macroexpansion. Like macroexpansion, also delegation allows for non-strict evaluation. This is described in section 9. As demonstrated in section 10 using arrays, delegation and non-strict evaluation allow a structured application definition to have a performant execution.

Previous dataflow models prohibit input/output, global items and other such effects. As described in section 11, TSIA places no restrictions on the effects of a task, provided that all effects are declared.

## 2 A Directed Acyclic Graph (DAG)

In the dataflow model, an executing application is represented as a directed acyclic graph (DAG). Each node of the graph denotes a task. Each arc of the graph denotes an item produced by one task and used by another.

In other presentations of the dataflow model, a task is named an actor [Models][Streams], a node [Architectures], a thread [Cilk-NOW][MIMD-Style][TAM], or another name. The names task and item are used here for consistency with previous presentations of TSIA.

A task consists of items: ins, inouts and outs. An in is an item used by the task. An inout is an item modified by the task. An out is an item produced by the task. An inout behaves like an in and an out. An item can be of arbitrary size and complexity.

One of the ins is the instruction of the task and represents the actions executed by the task. An instruction can be as small as a single machine instruction or as large as a million-line program. For convenience, an instruction is represented here as a routine. The syntax used is `routine(in,…;inout,…;out,…)`. The syntax is conventional, except that a semi-colon (`;`) separates the ins from the inouts and another semi-colon separates the inouts from the outs.

An example fragment of an executing application is given by the following graph consisting of three tasks.

```
plus(u,v;;a)
plus(v,w;;b)
mult(a,b;;c)
```
The items `u`, `v` and `w` are outs of some previous task or tasks. Similarly, `c` is an in of some subsequent task or tasks.

Since the result is undefined, an out of a task cannot also be another item of that task. For example, the task `plus(m,k;;k)` is undefined. In other words, aliasing involving outs is not permitted.

Instead of the above textual representation of a graph, a graphical illustration is usual in previous presentations of the dataflow model. For example, a graphical illustration of the above graph can be found elsewhere [Models]. Presentations of TSIA are served well by the textual representation.

In the textual representation of a graph, the name of an item is unique, but otherwise is arbitrary. As long as all occurrences are changed, an item may be arbitrarily renamed. For example, the above graph is equivalent to the graph `plus(u,v;;g) plus(v,w;;h) mult(g,h;;c)`.

An executing application is represented as a DAG in order to make explicit the dependencies between tasks. Previous dataflow models assume that explicit dependencies are incompatible with inouts [Architectures] [languages]. As demonstrated in this presentation, the dataflow model can support inouts. Among other benefits, inouts allow the modification of large data structures such as arrays.

An example using inouts is given by the following graph.
```
f(;x;)
g(;x;)
```
The task `f(;x;)` must execute before the task `g(;x;)`. In TSIA's textual representation of a graph, the dependency on an inout between tasks is top-to-bottom and left-to-right.

For convenience, the top-to-bottom and left-to-right order is used for all dependencies. A task producing an out is shown before any task which uses that item as an in. The order is suitable for a sequential execution. The order also is used in the programming language introduced later in this presentation.

## 3  An Execution in Terms of Tasks

In the dataflow model, an application executes in terms of tasks. A task executes to completion once its items are assembled. This defines a task.

For example, during its execution, a task does not communicate with other tasks.

The definition of a task allows each item of a task to be explicit. In turn, this allows the dependency on each item between tasks to be explicit. As introduced in the previous section, the dependencies between tasks are represented as a directed acyclic graph (DAG).

The graph is acyclic because the definition of a task does not allow cycles. No task in a cycle ever can execute. Since each task depends on others in the cycle, no task has the items required to execute.

The dependencies between tasks are obeyed by the dataflow system executing the application. Some consequences of the dataflow model are described below.

By definition, a task has no control over its own execution. By controlling the assembly of the items of a task, the dataflow system controls the execution of the task. For an application which executes in terms of tasks, the dataflow system thus controls the application execution.

A task may execute once its ins are available. If there are no dependencies between them, tasks can execute in any order, including in parallel. An example uses the graph of the previous section.

```
plus(u,v;;a) plus(v,w;;b) mult(a,b;;c)
```

If the items `u`, `v` and `w` are available then the tasks `plus(u,v;;a)` and `plus(v,w;;b)` can execute. Because there are no dependencies between them, the two tasks can execute in any order, including in parallel. In the dataflow model, the dependencies between tasks are explicit. Thus the available parallelism in an application execution is explicit.

In the above application execution, the resulting value for the item `c` is unaffected by the execution order of the tasks `plus(u,v;;a)` and `plus(v,w;;b)`. Because the dataflow model obeys the dependencies between tasks, an application execution has a determinate result.

## 4  The State of an Executing Application

As introduced in the previous sections, an executing application is represented as a directed acyclic graph (DAG). Because the graph is acyclic, a task executes once [Cilk-NOW][MIMD-Style][Monsoon][TAM]. This approach is pursued here. In contrast, some previous dataflow models introduced special items and tasks for cycles. Parts of the graph then may execute more than once in the application execution [Models][Architectures].

A graph conveniently represents the state of an executing application at a given instant. For example, the above example graph

    plus(u,v;;a) plus(v,w;;b) mult(a,b;;c)

represents an executing application whose state consists of the items `u`, `v`, `w`, `plus` and `mult`. The state includes the future execution of the application. The tasks `plus(u,v;;a) plus(v,w;;b) mult(a,b;;c)` are the future execution of this application.

Once a task has executed, it is removed from the graph, as are any items not needed by the remaining tasks. For example, after the execution of the task `plus(v,w;;b)`, the state of the above application execution is given by the graph

    plus(u,v;;a) mult(a,b;;c)

Since it has executed, the task `plus(v,w;;b)` no longer is part of the application state. Similarly, the item `w` no longer is part of the application state since it is not used by any of the remaining tasks.

In the dataflow model, the execution of a single task is the smallest unit of application execution. In other words, the execution of a single task is the smallest possible change to the state of an executing application. The execution of a task is considered to be indivisible. Thus the state within an executing task is not part of the dataflow model.

## 5 Delegation

In the example of the previous section, the execution of the task `plus(v,w;;b)` produced the out `b`. Instead of producing an out, the execution of a task can delegate the production to other tasks. In other words, a task in a graph can replace itself by other tasks. The eventual execution of the replacement tasks produces the out.

For example, the execution of a task `fact(1,3;;k)` might yield the tasks `fact(1,2;;x) fact(3,3;;y) mult(x,y;;k)`. In its execution, the task `fact(1,3;;k)` delegated the responsibility for its out `k` to the task `mult(x,y;;k)`. The eventual execution of the latter task requires the items `m` and `n` of the tasks `fact(1,2;;x)` and `fact(3,3;;y)`.

In many previous dataflow models, a task only can replace itself by a constant graph of tasks specified by the instruction of the original task [Architectures][Models][Monsoon][TAM]. This is called macroexpanding or copying the body of a routine.

An example of macroexpansion uses the routine

```
    add3(int a, int b, int c;; int d)
    { add(a,b;;r); add(r,c;;d); }
```
The programming language is introduced in the next section. Given the routine, the execution of a task `add3(6,9,17;;p)` results in the graph `add(6,9;;r) add(r,17;;p)`.

An advance over macroexpansion has been made by a recent dataflow model [Cilk-NOW][MIMD-Style]. A task can replace itself by any one of a variety of graphs, as determined by the ins of the original task. TSIA calls this technique delegation. It includes macroexpansion and is the approach pursued here. In delegation, the ins are used to execute the instruction and thus yield the replacement graph.

Macroexpansion yields many fine-grained tasks. One of the benefits of delegation is that it allows a coarser granularity. Adjusting the granularity can help application execution performance [Hybrid][Monsoon][TAM].

Delegation and its coarser granularity remain within the dataflow model and its tasks. In contrast, some other models for coarser granularity abandon the dataflow model. For example, a *scheduling quantum* is not a task [Hybrid]. A scheduling quantum does not execute to completion once its items are assembled.

An example of delegation uses the routines
```
    mult(int a, int b;; int c) { c = a*b; }

    // f = b * (b+1) * . . . * (e-1) * e
    fact(int b, int e;; int f)
    { if (b>=e) f=b;
      else      { int m=(b+e)/2;
                  fact(b,m;;x); fact(m+1,e;;y);
                  mult(x,y;;f);                  }
    }
```
The execution of a task `fact(1,3;;k)` yields the tasks `fact(1,2;;x) fact(3,3;;y) mult(x,y;;k)`. This is the example task and execution already used earlier in this section. Delegation and the example are further described in the next section.

## 6  A TSIA Language

In a dataflow model, each task is a black box. The implementation of a task thus is completely open. For example, the instruction of a task could be implemented in any programming language. TSIA thus is not restricted to a particular programming language. Instead, TSIA provides additional techniques to implement a programming language.

Following the real-world success of imperative languages and their extensions, this and previous presentations of TSIA use an extended imperative language. A further description of the language and many application examples are available elsewhere [TSIA]. Since the syntax is close to the TSIA semantics, the language is called a TSIA language. By choice, the imperative part of the language is similar to the C programming language.

The imperative part of the language is within a task. The other part is the dataflow part and is across tasks. Because a task has no hidden effects, the dataflow model leads to the benefits of an applicative language [TSIA]. The TSIA language of this presentation thus combines the benefits of an applicative language with those of an imperative language. An example of this combination is the use of non-strict evaluation to support arrays, as described in section 10.

The TSIA language of this presentation is designed to help clearly present TSIA and to demonstrate its feasibility. The language is not yet implemented. A similar TSIA language, with delegation but with few other features, has been implemented [Cilk-NOW][MIMD-Style].

For a routine written in a TSIA language, the execution proceeds as usual, except that calls to other routines yield the replacement graph of delegation. This alternative implementation of routines also is described in detail elsewhere [Alternative].

An example uses the routine `fact` and the task `fact(1,3;;k)` of the previous section. For the task, the routine executes as usual, except that the calls to the child tasks `fact(1,2;;x)`, `fact(3,3;;y)` and `mult(x,y;;k)` are not executed immediately. Instead, the child tasks are the replacement graph of delegation.

A child task thus does not return to its parent task. This is in contrast to the conventional implementation of routines.

Because a child task does not return to its parent, delegation forbids the parent task from using the outcome of its children. The routine `fact` is coded in this fashion. In contrast, the following otherwise equivalent routine is not.

```
    cfact(int b, int e;; int f)
   { if (b>=e) f=b;
      else      { int m=(b+e)/2;
                  cfact(b,m;;x); cfact(m+1,e;;y);
                  f=x*y;                          }
```

}

Any routine easily is mechanically translated into the delegation style. The offending code is replaced by calls to routines containing that code. In the above example, `f=x*y` is replaced by `mult(x,y;;f)`.

## 7 A Simple Execution

As introduced in the previous two sections, delegation is an alternative implementation of routines. It allows a simple application execution. The example execution demonstrated here is a sequential execution.

An executing application is represented as a directed acyclic graph (DAG). The tasks of the graph may be stored in a stack. The topmost task is executed and removed from the stack. This is repeated until the stack is empty, at which point the execution is complete.

An example execution uses the following application
```
    mult(int a, int b;; int c);
    fact(int b, int e;; int f);
    intprint(int i;;);
    main(;;) { fact(1,3;;k); intprint(k;;); }
```
The routines `mult` and `fact` are defined in section 5. The routine `intprint` outputs an integer onto some device.

Initially the stack is empty. An application execution is initiated by loading some task or tasks onto the stack. Here, `main(;;)` is loaded onto the stack. The initial state of the application execution thus is given by the graph
```
    main(;;)
```
Since it is the topmost task, `main(;;)` is executed and yields the graph
```
    fact(1,3;;k) intprint(k;;)
```
This is the next state of the application execution.

The stack contents may be illustrated as
```
    fact 1 3 &k intprint k
```
`fact` and `intprint` are references to these routines. `1` and `3` are `int` constants. `&k` is the address of the `int` `k` further down in the stack. So `k` is just a convenient name for an `int` in the stack. For simplicity, assume a routine receives the value of an in and the reference to an out.

Execution of the topmost task yields the graph
```
    fact(1,2;;x) fact(3,3;;y) mult(x,y;;k)
    intprint(k;;)
```
Execution of the topmost task yields

```
    fact(1,1;;a) fact(2,2;;b) mult(a,b;;x)
    fact(3,3;;y) mult(x,y;;k)
    intprint(k;;)
```
The items `a` and `b` have been renamed for clarity.

Execution of the topmost task writes the value `1` to `a` on the stack. Similarly, `b=2` from the next topmost task. So, the graph now is
```
    mult(1,2;;x) fact(3,3;;y) mult(x,y;;k)
    intprint(k;;)
```
Similarly, `x=2` and `y=3` from the two next topmost tasks. So, the graph now is
```
    mult(2,3;;k) intprint(k;;)
```
Execution of the topmost task yields the graph
```
    intprint(6;;)
```
Execution of the topmost task outputs `6` onto some device. Since the stack is empty, the execution is complete.

Cilk-NOW and its immediate precursors are dataflow systems which implement an execution very similar to that demonstrated above [Cilk-NOW][MIMD-Style]. Their extension from the above sequential execution to a parallel execution is simple. Each computer has its own stack of tasks. By always executing the topmost task, each computer performs a depth-first execution as demonstrated above. When a computer's stack is empty, it steals a ready-to-execute task from the bottom of another computer's stack. This breadth-first execution thus only is performed when required.

The particular sequential execution demonstrated above for the alternative implementation of routines is briefly compared below to that of the conventional implementation of routines. After forty years, the conventional implementation is still well described by its original proposal [Stack]. In the execution of the conventional implementation, a parent routine waits for its child to return. The parent can continue its execution, since its state is stored on the stack during the execution of the child.

The conventional implementation uses a program counter to step through the execution. In contrast, no program counter is used by the alternative implementation since it always executes the topmost task on the stack.

For the conventional implementation, the stack stores the past execution of the application. For example, a stack trace shows the chain of called routines. In contrast, the stack of the alternative implementation stores the future execution of the application.

The stack of the conventional implementation may contain items irrelevant to the future application execution. In contrast, the stack of the alternative implementation contains only items used in the future application execution. This feature is named proper tail calls [Alternative]. For other execution features, such as heterogeneity and reliability, the advantages of the alternative implementation are described elsewhere [TSIA].

As mentioned in the previous section, a task is a black box. Thus the implementation of a task thus is completely open. So within a task may well be the conventional implementation of routines. For example, the call to a routine may use the alternative or the conventional implementation. An application thus may execute as many fine-grained tasks or as few coarse-grained tasks [TSIA]. Adjusting the granularity can help execution performance [Hybrid][Monsoon][TAM].

## 8  No Special Items

As demonstrated in the example execution of the previous section, TSIA blindly executes tasks, with no regard for any special instructions nor for any other special items. Like some other dataflow models, TSIA has no special items [Cilk-NOW][MIMD-Style].

In contrast, special items are part of the semantics of some dataflow models. Special actors are examples of special items [Models].

Since TSIA contains no special items, its semantics do not include the semantics of any item. The semantics of an item are part of the application, not part of the TSIA.

## 9  Non-Strict Evaluation

The macroexpansion of previous dataflow models allows for non-strict evaluation [Models]. Delegation is more general than macroexpansion. This section shows that delegation also allows for non-strict evaluation. This section largely is taken from elsewhere [Alternative].

A strict evaluation requires the argument of a routine to be evaluated before the execution of the routine. Most of current conventional computing involves strict evaluation.

A TSIA can provide strict evaluation. An example is given by the graph `fact(1,2;;x) fact(3,3;;y) mult(x,y;;k)` in the section 7. Only after `fact(1,2;;x)` has evaluated `x` and `fact(3,3;;y)` has evaluated `y`, can `mult(x,y;;k)` execute. The task `mult(x,y;;k)` requires a strict evaluation for the items `x` and `y`.

In a non-strict evaluation, the evaluation of the argument of a routine is not restricted to occur before the execution of the routine. Instead, the argument may be evaluated before, during or after the execution of the routine.

Delegation allows for a simple implementation of non-strict evaluation. In the TSIA language of this presentation, the implementation uses the keyword `del` to declare that an item of a task is delegated. This overrides the default assumption that an item is evaluated by its task and thus requires a strict evaluation.

For example in Figure 1a), the routine `b(p;del x;)` delegates the item `x` to the task `c(;x;)` or to the tasks `c(;x;) c(;x;)`, depending on `p`. The routine thus does not require a strict evaluation of `x`. In contrast, the routine does require a strict evaluation of `p`.

As demonstrated above, the keyword `del` describes a property of an item of a routine. The property can be derived from the definition of the routine. The keyword `del` is not an arbitrary annotation. For example, the routine `b(p;del x;)` would require rewriting to be not strict in `p`.

Because delegation affects the allowed executions, presentations of tasks and graphs may conveniently include the keyword `del` where appropriate. This is demonstrated in the following text and in Figure 1b) through d).

Non-strict evaluation in the TSIA can be demonstrated using the code of Figure 1a). The prototypes `a(;;int x)` and `c(;int x;)` are declarations like in the C programming language. The definitions of these routines are not of interest here. Since the routine `b(p;del x;)` delegates `x` to `c(;x;)`, it does not require a strict evaluation of `x`. For example, if a graph originally consists of the tasks `a(;;q) b(false; del q;)`, then non-strict evaluation allows for the any of the three different executions illustrated in Figure 1b), c) and d). In Figure 1b), `a(;;q)` evaluates `q` before the execution of `b(false;del q;)`, as in a strict evaluation. In Figure 1c), `a(;;q)` evaluates `q` after the execution of `b(false;del q;)`. In Figure 1d), `a(;;q)` evaluates `q` in parallel during the execution of `b(false;del q;)`.

In the illustrations of graph execution in Figure 1b), c) and d), an arrow illustrates the execution of a task. A line illustrates a task remaining as is in the graph.

Evaluation in macroexpansion and in delegation now can be compared. In macroexpansion, the instruction is the only strict in, since only the instruction is used to yield the replacement graph. All other ins are non-strict. In other words, all arguments of the instruction are non-strict.

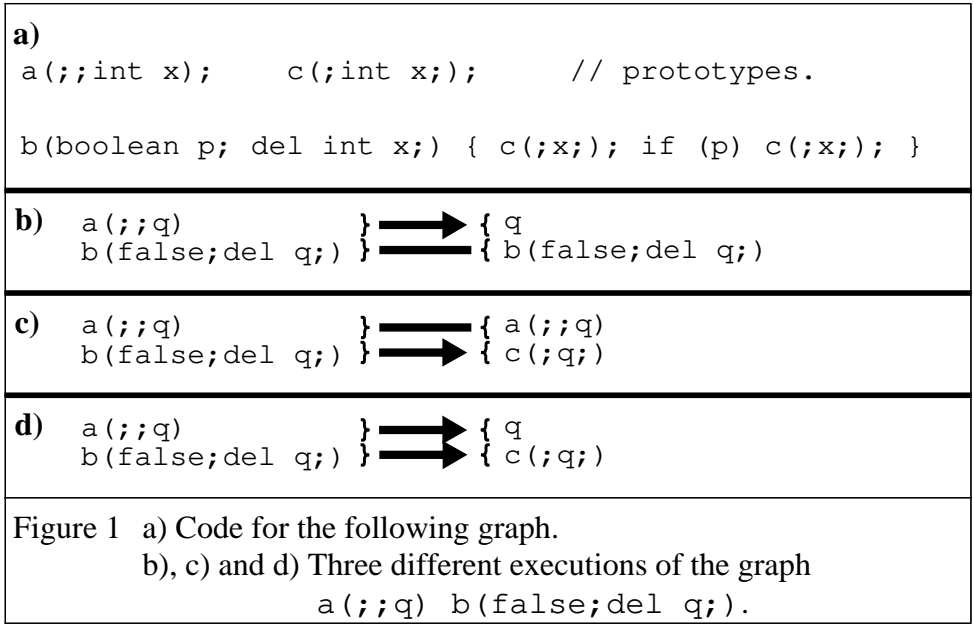

Figure 1  a) Code for the following graph.
    b), c) and d) Three different executions of the graph
    `a(;;q) b(false;del q;)`.

In delegation, the instruction is strict as may be some other ins, since the instruction and those other ins are used to yield the replacement graph. The remaining ins are non-strict. In other words, some arguments of the instruction may be non-strict.

Non-strict evaluation allows a variety of language features. An example is streams, using macroexpansion [Streams] or delegation [Alternative] [TSIA]. Another example is conditional items [Alternative].

Non-strict evaluation also is valuable since it allows for a variety of executions. One of the executions may make better use of resources or otherwise may better meet the execution requirements of the application. This is demonstrated for arrays in the next section.

## 10  Arrays

For an application involving large data structures such as arrays, a number of issues must be satisfied in order to achieve a performant execution. Two such issues are addressed in this section. These and other issues are discussed in more detail elsewhere [TSIA].

The first issue requires that the elements of an array can be modified. As introduced in section 2, TSIA supports inouts.

In contrast, previous dataflow models assume inouts to be incompatible with the model. Thus any modification of an array, even to just a single ele-ment of the array, requires creating a new array. The new array is largely a copy of the old array. Such copying is detrimental to a performant execution. For example, implementations of previous dataflow models use an optimization called copy elimination [SISAL]. In retrospect, this indirect support for inouts by the implementations contradicts the assumption against inouts by their models.

TSIA also allows the copying mentioned above. For example, the graph `a(x;;) b(;x;)` can have an execution which copies `x` in order that the two tasks can execute in parallel.

The second issue addressed here is the required flexibility for placing the elements of an array. A performant execution minimizes communication by placing related items together in space and/or time. A good placement is known as good locality.

An example uses the following toy application.

```
// Definitions and declarations of routines.
b(;; int y[0:4999]);
c(;; int y[0:4999]);
a(;; del int y[0:9999])
   { b(;;y[0]); c(;;y[5000]); }

d(; int y[0:4999];);
e(; del int y[0:9999];)
   { d(;;y); d(;;y[5000]); }

// Application fragment.
a(;;h);
e(;h;);
```

The TSIA language of this presentation allows unambiguous shorthand. For example, using the definition of its routine, `a(;;h)` is shorthand for `a(;;h[0:9999])` and thus creates an array of 10000 elements.

TSIA provides the required flexibility for placing the elements of an array. The flexibility is a result of delegation. If a task delegates an item, then the task only needs a reference or pointer to that item. The task does not need the delegated item itself. In the above example, the execution of `c(;h;)` does not require array `h`. The task `c(;h;)` can execute on a computer, with array `h` distributed on many computers. In fact, no part of the array `h` need be on the computer executing `c(;h;)`. Because of delegation, the placement of the array `h` is flexible.

In contrast, without delegation there is no such flexibility. If a task evaluates an item, then it needs the item itself. In the above example, the execution of `d(;;y[0:4999])` requires all 5000 elements of that (sub)array on the computer.

The benefit of flexibility is illustrated in Figure 2 using a two computer parallel execution of the above application. The task `a(;;del h)` exe-cutes on either computer 1 or 2. The resulting tasks `b(;;h[0:4999])` and `c(;;h[5000:9999])` are assumed to execute on computers 1 and 2, respectively. The subarrays `h[0:4999]` and `h[5000:9999]` are on computers 1 and 2, respectively. Similarly, the task `e(;del h;)` executes on either computer 1 or 2. The resulting tasks `d(;;h[0:4999])` and `d(;;h[5000:9999])` are assumed to execute on computers 1 and 2, respectively. The assumed execution avoids moving the array. This good locality is possible due to the flexibility provided by delegation.

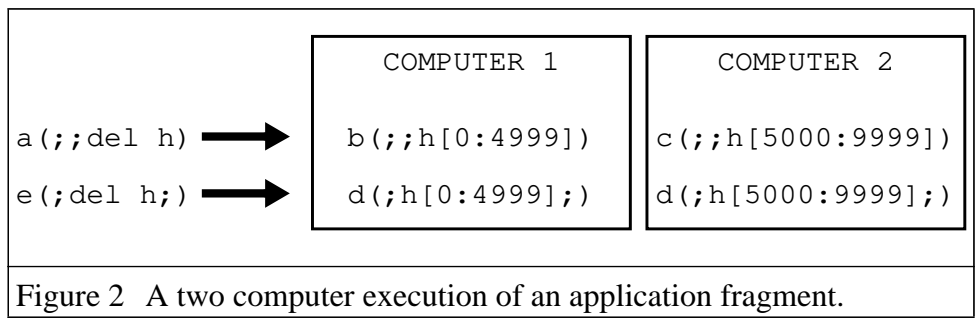

Figure 2  A two computer execution of an application fragment.

There is a large distinction between providing good locality and providing the required flexibility. As a model for transparent application execution, TSIA does not magically provide good locality. The TSIA model merely provides the required flexibility. Achieving good locality is part of the efforts of a system implementing TSIA. In general, a model for transparent application execution does not magically provide a performant execution. Instead, the model merely moves the execution issues from the application to some external system. This implies that the system is given a sufficiently flexible definition of the application; a definition which allows the system to provide a performant execution.

Without delegation, the equivalent execution flexibility requires a less structured application definition. For example, instead of calling a routine, the body of the routine is copied into the definition. For the above example, the original definition `a(;;h);e(;h;)` would be replaced by the less structured `b(;;h);c(;;h[5000]);d(;h;);d(;h[5000];)`. In general, delegation and non-strict evaluation allows a structured application definition to have a performant execution [TSIA].

## 11 Input/Output, Global Items and Other Such Effects

Previous dataflow models prohibit most effects.
> "...a data flow computer imposes much stricter prohibitions against side effects--a procedure may not even modify its own arguments. In fact, in a sense nothing may ever be modified at all." [languages]

In previous dataflow models, the effect of a task is the production of a value.

In contrast, TSIA places no restrictions on the effects of task, provided that all effects are declared. The dependencies between tasks thus are explicit and can be obeyed by a TSIA executing the application.

Up until this point of the presentation, a task only has local effects. Such effects involve local items defined by the parent of the task. Nonlocal effects involve nonlocal items such as global items or involve input, output or other such effects.

The TSIA language of this presentation takes from elsewhere the declaration of nonlocal effects [Scope]. There it has other motivations. For example, a routine to print a character on standard output is declared as

```
putc(char i;;)(;stdout;);
```

The first set of parentheses declares the local items. The second set declares the nonlocal items.

Because a nonlocal item affects the allowed executions, presentations of tasks and graphs may conveniently include the nonlocal item where appropriate. For example the application definition `putc('1';;); putc('2';;)` corresponds to the graph

```
putc('1';;)(;stdout;) putc('2';;)(;stdout;)
```

The graph thus makes explicit that the two tasks must execute in order, since each modifies `stdout`.

A slightly more involved execution is demonstrated by the following application.

```
putc(char i;;)(;stdout;);
puts(char s[];;)
{ if (s[0]!=0) { putc(s[0];;);
                 puts(s[1];;); }
}
puts("AB";;); puts("CD";;);
```

The original graph of the executing application is

```
    puts("AB";;) puts("CD";;)
```
A `puts` task does not modify `stdout`. Instead, it merely calls `putc` which modifies `stdout`. The `puts` tasks thus can execute in any order, including in parallel.

In the execution illustrated here, `puts("CD";;)` is assumed to execute first. This yields the graph
```
    puts("AB";;)
    putc('C';;)(;stdout;)
    putc('D';;)(;stdout;)
```
At this point in the execution, `putc('D';;)(;stdout;)` cannot execute yet because `stdout` is to be modified by the prior task `putc('C';;)(;stdout;)`. Similarly, `putc('C';;)(;stdout;)` cannot execute yet because `puts("AB";;)` or any other prior tasks may delegate to a child task which modifies `stdout`.

The execution of the task `puts("AB";;)` yields the graph
```
    putc('A';;)(;stdout;)
    putc('B';;)(;stdout;)
    putc('C';;)(;stdout;)
    putc('D';;)(;stdout;)
```
Since there are no prior tasks, `putc('A';;)(;stdout;)` may execute and print the character A on standard output. In turn, the characters B, C and D also are printed to standard output. Thus as required by the application definition, the characters ABCD are printed to standard output in the defined order.

In an alternative implementation of TSIA, the above routine `puts` could be declared as `puts(char s[];;)(;del stdout;)`. The declaration of `puts` thus includes the call to `putc` which modifies `stdout`. This might be considered to be a more complete declaration of nonlocal effects.

The more complete declaration can be used in the graph of an executing application. For example, the original graph of the above application then would be
```
    puts("AB";;)(;del stdout;)
    puts("CD";;)(;del stdout;)
```
As above, the `puts` tasks thus can execute in any order, including in parallel. Assuming again that `puts("CD";;)(;del stdout;)` executes first yields the graph
```
    puts("AB";;)(;del stdout;)
    putc('C';;)(;stdout;)
```

```
    putc('D';;)(;stdout;)
```
The task `putc('C';;)(;stdout;)` cannot execute yet because the prior task `puts("AB";;)(;del stdout;)` delegates to a child task which modifies `stdout`. The execution of `putc('C';;)(;stdout;)` can ignore any prior task not involving `stdout`. This might help improve the performance of the application execution.

Any routine calling `puts` also would declare the nonlocal effect `(;del stdout;)`. An example is the routine `putint` shown here.
```
    int2chars(int i;;char s[]);
    puts(char s[];;)(;del stdout;);
    putint(int i;;)(;del stdout;)
    { int2chars(i;;s); puts(s;;); }
```
In turn, any routine calling `putint` also would declare the nonlocal effect `(;del stdout;)`. In general, the declarations of nonlocal effects would have to be propagated up the call chain of the application definition. Such propagation is tedious and error-prone for a person. Thus the propagation presumably would be performed by the TSIA implementation.

Without delegation, the more complete declaration for the routine `puts` would be
```
    puts(char s[];;)(;stdout;);
```
Without delegation, the execution of `puts` would lose much flexibility. A `puts` task would have an execution like `putc`, since its declaration of `stdout` is like that of `putc`. Before such a `puts` task could execute, there could be no prior tasks which involve `stdout`. For example, `puts` tasks could not execute in parallel.

Without delegation, the more complete declaration of a routine seems to be similar to the declaration of nonlocal effects in a functional language using monads. There is an introduction to monads which includes the above `putc` and `puts` example using the programming language Haskell [Scope]. The declarations of the routines are repeated here.
```
    putc :: Char   -> IO ()
    puts :: String -> IO ()
```
In Haskell with monads, `putc` and `puts` have the same declaration of nonlocal effects, even though `puts` is implemented in terms of `putc`.

## 12 Conclusion

TSIA is one of a succession of dataflow models. With TSIA, a simple and powerful dataflow model is achieved.

In many real-world projects, a TSIA provides a simple application with a transparent reliable, distributed, heterogeneous, adaptive, dynamic, realtime, parallel, secure or other execution [Dividing]. With its simple and powerful model, TSIA is suitable for many applications, not just for the simple applications served to date.

**Acknowledgments**

Previous presentations of TSIA led several people to ask about the relationship between TSIA and dataflow models. This presentation is a response to those questions.